\documentclass[12pt]{article}
\usepackage[textwidth=18cm,textheight=24cm]{geometry}
\usepackage{graphicx}

\newtheorem{Th}{Theorem}
\begin{document}

\title{Analytic calibration in Andreasen-Huge SABR model.}

\author{Konstantin Feldman\footnote{This paper is a personal view and does not represent the views of MUFG Securities EMA plc (“MUSE”). This paper is not advice.  MUSE shall not be liable in any manner whatsoever for any consequences or loss (including but not limited to any direct, indirect or consequential loss, loss of profits and damages)
 arising from any reliance on or usage of this presentation and accepts no legal responsibility to any party who directly or indirectly receives this material.}}
\date{MUFG Securities EMEA plc.}
\maketitle{}

\begin{abstract}
We derive analytic formulae which link $\alpha$, $\nu$ and $\rho$ parameters in Andreasen-Huge style SABR model to the ATM price and option prices at four  strikes close to ATM. Based on these formulae we give a characterisation for the SABR parameters in terms of derivatives of the swap rate forward probability density function. We test the analytic result in the application to the interest rate futures option market. 
\end{abstract}

\section{Introduction}
20 years ago Patrik Hagan, Deep Kumar, Andrew Lesniewski and Diana Woodward published a paper~\cite{HKLW1} which revolutionised interest rate volatility modelling. They derived a closed form analytic approximation for the implied volatility of a swap rate forward assuming it satisfies a  system of stochastic differential equations:
\begin{eqnarray}
dF_t &=& \alpha_t F_t^{\beta} d W_t;\nonumber\\
d\alpha_t &=& \nu \alpha_t d Z_t;\nonumber\\
\rho &=& <W_t, Z_t >.
\label{SABRSDE}
\end{eqnarray}
The analytic approximation gave interest rate traders a simple tool to manage the swap rate probability density distributions which are skewed and smiled over the traditional Gaussian normal distribution. The model got a name  - SABR model.

Various  improvements were made to the analytical formula from~\cite{HKLW1} in works~\cite{BBF,L} and even more recently in~\cite{AKS}.

One of the issues which became transparent in the industrial use of SABR model was its approximation character. As it was an approximation for the forwards behavior around the ATM level, it couldn't possibly give the terminal swap rate probability density at every point. Polynomial expansions used  in the formula were unbounded at infinity and, thus, led to negative values of the implied density function. 

A striking approach to address the problem of constructing terminal swap rate probability density valid  along the whole real line appeared 10 years later in another seminal paper~\cite{AH1}. Jesper Andreasen and Brian Huge showed how within SABR model assumptions one can reduce the construction of the swap rate forward probability density function to a solution of a single tridiagonal system.
The construction not only resolved the issue of the validity of the probability density function at every point, but also prompted more widespread use of the density functions for modelling light exotics like CMSs and CMS spreads, which were often done by a slower replication based methods.  While some market participants went for the approach~\cite{AH1}, other developments also made into production systems of many banks~\cite{BT,HKLW2,B}.   

An important technical problem which accompanies the use of SABR model is the parameters' calibration. The parameter $\beta$ can be estimated from the log-log plot of the forward variance with respect to the  forward~\cite{HKLW1} and $\alpha$ could be found as a root of a cubic equation~\cite{W}. The skew parameters $\nu$ and $\rho$ were often manually calibrated. Multi dimensional solvers were introduced to address parameter mapping while migrating to the shifted SABR model which became a necessity as the rates entered negative territory. Extensive studies have been made on how to chose the initial guess to optimise solver convergence ~\cite{LFK}.

Andreasen-Huge construction of the non-arbitrage SABR did not contain explicit analytics to invert, and, thus, required use of solvers to imply not only the skew parameters, but the parameter alpha as well. As Andreasen-Huge construction gives considerable computational speed up in comparison with the other arbitrage free SABR methods, the use of solvers did not represent a limitation for building infrastructure around their version of SABR model. The goal of the present paper is to show that Andreasen-Huge construction of SABR model, in fact, allows one to have exact analytic expressions of all three model parameters: alpha, nu and rho, in terms of option prices close to ATM. In the paper, we show that if call prices $c(F-2h),c(F-h), c(F), c(F+h), c(F+2h)$ are given at five consecutive grid points of an Andresen-Huge shifted SABR grid with $c(F)=ATM$ - the ATM option price, then: 
\begin{eqnarray}
\alpha &=& \frac{h}{(F+b)^{\beta}}\sqrt{\frac{1}{T}}\sqrt{\frac{ATM }{c(F-h) + c(F+h) - 2ATM}},\\
\nu^2 & = &\frac{\frac{z_{-}h^2}{ y(F-h)  (F - h + b)^{2\beta}}
- \frac{z_{+}h^2}{  y(F+h)(F+h + b)^{2\beta}}}{T\kappa_h\alpha^2(y(F-h) - y(F+h))} - \frac{1}{y(F-h) y(F+h)},\\
\rho &=& \frac{1}{2\nu}\left(\nu^2 y(F-h) - \frac{1}{y(F-h)}\left(\frac{z_{-}h^2}{T \kappa_h\alpha^2   (F-h + b)^{2\beta}}-1\right)\right),
\end{eqnarray}
where $y(k)$ is the shifted SABR local volatility diffusion distance from the forward to the strike, $\kappa_h$ is the one half of the value of the Andreasen-Huge local volatility one time step ODE adjustment at the strikes $F\pm h$ and 
\begin{eqnarray}
z_{-} =  \frac{c(F-h) - h}{ c(F-2h) +ATM - 2c(F-h) },& \quad{}&
z_{+}=  \frac{c(F+h)}{ c(F+2h) + ATM - 2c(F+h)}.
\end{eqnarray}
Note, that in the paper we derive the analytic formulae not just for a uniform grid but for any types of grids. In the case of a uniform grid the formulae are  more compact as they don't need to include a reference to a varying size of the grid steps.  

Using the analytic expressions we study the limiting behavior when $h\to 0$, and derive a characterisation for SABR parameters in terms of probability density function of the swap rate forward distribution. We discuss two applications. 1) We show how these formulae can be used for migrating from/between analytic expansions~\cite{HKLW1} and the Andreasen-Huge construction. 2) We show how analytic formulas for SABR parameters perform in EURODOLLAR option market where the options strikes are quoted with a uniform interval of a sensible local volatility grid. 

The structure of the paper is as follows. In Section 1 we remind the Andreasen-Huge construction for SABR model. In Section 2 we derive analytic formulae for (shifted) SABR parameters in the Andreasen-Huge construction for a non-uniform grid. In Section 3 we study the limiting behavior of the analytic formulae. In Section 4 we  show how the Andreasen-Huge construction can be calibrated to the Hagan etl. original construction. In Section 5 we test the performance of the analytic expressions in the EURODOLLAR option market. We summarise our findings in the conclusion.

\section{Andreasen-Huge one-time-step method}

In this section we remind the Andreasen-Huge one time step ODE reduction for SABR SDE~\cite{AH1}.
 It rests on their other remarkable paper~\cite{AH2} where they showed that the option pricing heat equation:
\begin{eqnarray}
c_t(t,k) &=& \frac{1}{2}\vartheta(k)^2 c_{kk}(t,k)\nonumber\\
c(T,k)& =&(F-k)^+,
\end{eqnarray}
 with the local volatility coefficient $\vartheta(k)$ dependent only on the strike $k$ and the terminal condition given by the intrinsic value of a call option, can be replaced by an ODE:
\begin{eqnarray}
c(t,k)-\frac{1}{2}(T-t)\theta(k)^2c_{kk}(t,k) & = & (F-k)^+.
\label{AHODE}
\end{eqnarray}
Based on this reduction, Andreasen and Huge derived~\cite{AH1} the relation between $\theta(k)$ and $\vartheta(k)$ in terms of the standard normal distribution:
\begin{eqnarray}
\theta(k)^2 &=& \vartheta(k)^2 \frac{c(t,k)-(F-k)^+}{(T-t)c_t(t,k)},\nonumber\\
&=& 2\vartheta(k)^2\left(1-\xi\frac{\Phi(-\xi)}{\phi(\xi)}\right),  
\label{REODE}
\end{eqnarray} 
where $\Phi(\xi)$ is the CDF of the standard normal random variable, $\phi(\xi)$ is the PDF of the standard normal random variable and $\xi=|F-k|/\sigma$, with $\sigma$ being ATM normal volatility. In what follows we shall be calling the coefficient
\begin{eqnarray}
\kappa(k) &=& 2\left(1-\xi\frac{\Phi(-\xi)}{\phi(\xi)}\right)  
\end{eqnarray}
as Andreasen-Huge local volatility one time step adjustment.

The ODE~(\ref{AHODE}) is solved by inverting a tridiagonal matrix corresponding to the finite difference equation~\cite{AH2}:
\begin{eqnarray}
\left(1 + \frac{T\theta(k)^2}{h_+h_-}\right) c(0,k) -  \frac{T\theta(k)^2}{h_++h_-}\left[\frac{c(0,k+h_+)}{h_+} + \frac{c(0,k-h_-)}{h_-}\right] = (F-k)^+.
\label{tridc}
\end{eqnarray}
In the identical way we can solve the option pricing problem in terms of puts:
\begin{eqnarray}
\left(1 + \frac{T\theta(k)^2}{h_+h_-}\right) p(0,k) -  \frac{T\theta(k)^2}{h_++h_-}\left[\frac{p(0,k+h_+)}{h_+} + \frac{p(0,k-h_-)}{h_-}\right] = (k-F)^+.
\label{tridp}
\end{eqnarray}
Note, that when reducing~(\ref{tridc}) or~(\ref{tridp}) to solving the linear system with a tridiagonal matrix we need to set an absorbing boundary condition~\cite{AH2}:
\begin{eqnarray}
c_{kk}(0,k_{-\infty}) &= c_{kk}(0,k_{\infty}) &=0,\\
p_{kk}(0,k_{-\infty}) &= p_{kk}(0,k_{\infty}) &=0.
\end{eqnarray}

In Andreasen-Huge SABR construction~\cite{AH1} we capitalise on the analytic form of the local volatility $\vartheta(k)$ for SABR SDE~(\ref{SABRSDE}).
Using shifted version of the SABR SDE~(\ref{SABRSDE}) we can write~\cite{AH1}:
 \begin{eqnarray}
\label{ahd1}
\vartheta(k) &=& \alpha J(y) (k+b)^{\beta},\\
y(k)&=&\frac{1}{\alpha}\int^F_k (u+b)^{-\beta}du,\\
\label{ahd2}
J(y) & = & \sqrt{1 - 2\rho \nu y + \nu^2 y^2}.
\label{ahd3}
\end{eqnarray}

Thus, in order to solve for option prices in (shifted) SABR SDE~(\ref{SABRSDE}) we can solve linear tridiagonal system~(\ref{tridc}) or~(\ref{tridp}), whose coefficients are populated using reduction~(\ref{REODE}) and analytic formulae~(\ref{ahd1}-\ref{ahd3}). 

In  further sections, we will need explicitly the values of the local volatility one time step adjustments for $k= F-h, F, F+h$  respectively, where $F$ is the forward and $h$ is the step of the uniform grid in~(\ref{tridc}) (or~(\ref{tridp})). They can be calculated as below:
\begin{eqnarray}
\kappa(F) & =  & 2,\\
\kappa(F\pm h) &=& 2 \left(1 -  \frac{h}{\sigma\phi(h/\sigma)}\left(\frac{1}{2} - \int^{h/\sigma}_0\phi(s)ds\right) \right)\\
&=& 2 \left(1 -  \frac{h}{\sigma\phi(h/\sigma)}\left( \frac{1}{2} - \frac{h}{\sigma}\phi(\zeta) \right) \right), \quad{} \zeta\in [0;h/\sigma].
\end{eqnarray}
In what follows we shall be using a notation $\kappa_h = \kappa(F\pm h)/2$.

\section{Analytic calibration}
In this section we use Andreasen-Huge one time step ODE reduction of the  (shifted) SABR SDE to derive exact formulae which relate SABR SDE coefficients $\alpha$, $\nu$ and $\rho$ to the prices of five options close to ATM. In what follows we shall assume that the finite difference grid for~(\ref{tridc},\ref{tridp}) contains the swap rate forward as one of the grid points.

First we write the  the one time step approach of the previous section explicitly in terms of put option prices~(\ref{tridp}).
The values of all OTM puts $p_0, \dots\ p_{n-1}$, ATM put $p_n$ and the first ITM put $p_{n+1}$  satisfy the following tridiagonal system:
\[
\label{redputsys}
 \left(\begin{array}{@{}cccccccccc@{}}
1 & 0      &   0          	  &0     & \cdots                               &0  &0         	       & 0                  & 0 \\
\frac{-z_1h^+_{1}}{h^+_{1}+h^-_{1}}      & 1+z_1 & \frac{-z_1h^-_{1}}{h^+_{1} +h^-_{1}}            &0   & \cdots    &0                            & 0         	       & 0                  & 0\\
0& \frac{-z_2h^+_{2}}{h^+_{2}+h^-_{2}}      & 1+z_2        &\frac{-z_2h^-_{2}}{h^+_{2} +h^-_{2}}        & \cdots    &0                            & 0         	       & 0                  & 0\\

\vdots  & \vdots   & \vdots   & \vdots   & \vdots         &\vdots & \vdots 	       & \vdots          \\
0          & 0           & 0            &0       & \cdots                       & 1+ z_{n-2}     & \frac{-z_{n-2}h^-_{n-2}}{h^+_{n-2}+h^-_{n-2}}  &0      & 0\\
 
 0          & 0           & 0            &0       & \cdots              & \frac{-z_{n-1}h^+_{n-1}}{h^+_{n-1}+h^-_{n-1}}         & 1+ z_{n-1}     & \frac{-z_{n-1}h^-_{n-1}}{h^+_{n-1}+h^-_{n-1}}        & 0\\
 0          & 0           & 0                &0   & \cdots              	        &0   & \frac{-z_nh^+_{n}}{h^+_{n}+h^-_{n}}                & 1+ z_n         & \frac{-z_nh^-_{n}}{h^+_{n}+h^-_{n}}\\
                  \end{array}\right) \\[15pt]
 \left(\begin{array}{@{}c@{}}
p_0 \\
p_1     \\
p_2       \\
p_3       \\
\vdots \\
p_{n-2}    \\
 p_{n-1}    \\
 p_n  \\
p_{n+1}  \\
                  \end{array}\right) \\[15pt]=
 \left(\begin{array}{@{}c@{}}
0 \\
0     \\
0       \\
0       \\
\vdots \\
 0    \\
 0    \\
 0  \\
0  \\
                  \end{array}\right) \\[15pt],
\]
where the coefficients $z_j$, $j = 0,\dots n$, are given by~(\ref{tridp}):
\begin{eqnarray}
z_j & = &  \frac{T\theta(k_j)^2}{h^+_jh^-_j}, \quad{} h^+_j = k_{j+1} - k_j, \quad h^-_j = k_{j} - k_{j-1}.
\end{eqnarray}
The value $ATM=p_n$ is (typically) observed from the market and is 
$$
ATM=p_n=\sqrt{\frac{T}{2\pi}}\sigma_{ATM},
$$
whith $\sigma_{ATM}$ - the implied normal volatility of the ATM option (put or call).

 Note that the tridiagonal system above is a sub system of the full tridiagonal system required to solve for option prices in the Andreasen-Huge method. We look only into the first $n+1$-equation on the first $(n+2)$-put prices  and equate them to the intrinsic values of all ITM puts and the ATM put - all these values are zeros. 

We can reduce the constructed tridiagoanl system by one dimension to
\[
 \left(\begin{array}{@{}cccccccccc@{}}
1 & 0      &   0     &0     	       & \cdots              & 0                  & 0         	                         & 0 \\
\frac{-z_1h^+_{1}}{h^-_{1}+h^+_{1}}      & 1+z_1 & \frac{-z_1h^-_{1}}{h^-_{1}+h^+_{1}}    &0           & \cdots               & 0                  & 0         	                         & 0\\
0 & \frac{-z_2h^+_{2}}{h^-_{2}+h^+_{2}}      & 1+z_2      & \frac{-z_2h^-_{2}}{h^-_{2}+h^+_{2}}          & \cdots               & 0                  & 0         	                         & 0\\
\vdots  & \vdots   & \vdots & \vdots   & \vdots & \vdots            & \vdots 	       	              & \vdots\\
 0          & 0           & 0     &0              & \cdots  & 1+ z_{n-2} & \frac{-z_{n-2}h^-_{n-2}}{h^-_{n-2}+h^+_{n-2}}                              & 0\\
 0          & 0           & 0     &0             & \cdots              & \frac{-z_{n-1}h^+_{n-1}}{h^-_{n-1}+h^+_{n-1}}       & 1+ z_{n-1}                      & 0\\
 0          & 0           & 0     &0             & \cdots              &  0	           & \frac{-z_nh^+_{n}}{h^-_{n}+h^+_{n}}                                   & \frac{-z_nh^-_{n}}{h^-_{n}+h^+_{n}}\\
                  \end{array}\right) 
 \left(\begin{array}{@{}c@{}}
p_0 \\
p_1     \\
p_2     \\
p_3     \\
\vdots \\
p_{n-2}       \\
 p_{n-1}    \\
 p_{n+1}  \\
                  \end{array}\right) \\[12pt]=
 \left(\begin{array}{@{}c@{}}
0 \\
0     \\
0       \\
0       \\
\vdots \\
 0    \\
 \frac{z_{n-1}h^-_{n-1}}{h^-_{n-1}+h^+_{n-1}}ATM  \\
 - (1+z_n)ATM  \\
                  \end{array}\right) \\[12pt].
\]
%0           & -z_2      & 1+2z_2 &  -z_2   & \cdots  & 0             & 0                  & 0         	                         & 0\\
The last equation  of the system:
\begin{eqnarray}
z_n(p_{n-1}h^-_{n} + p_{n+1}h^+_{n}) &=& (1+z_n)ATM(h^+_{n}+h^-_{n})
\label{atmterm}
\end{eqnarray}
 can be solved separately after solving 
\[
 \left(\begin{array}{@{}cccccccccc@{}}
1 & 0      &   0         	&0       & \cdots               & 0                 & 0 & 0         	                      \\
\frac{-z_1h^+_1}{h^+_{1}+h^-_{1}}      & 1+z_1 & \frac{-z_1h^-_1}{h^+_{1}+h^-_{1}}  &0            & \cdots         & 0      & 0                  & 0         	                      \\
0 & \frac{-z_2h^+_{2}}{h^-_{2}+h^+_{2}}      & 1+z_2      & \frac{-z_2h^-_{2}}{h^-_{2}+h^+_{2}}          & \cdots & 0               & 0                  & 0         	                         \\
\vdots  & \vdots   & \vdots    & \vdots & \vdots  & \vdots & \vdots     & \vdots            	       	           \\
 0          & 0           & 0      &0            &\cdots & \frac{-z_{n-2}h^+_{n-2}}{h^+_{n-2}+h^-_{n-2}} & 1+ z_{n-2} & \frac{-z_{n-2}h^-_{n-2}}{h^+_{n-2}+h^-_{n-2}}                          \\
 0          & 0           & 0       &0            & \cdots        & 0       & \frac{-z_{n-1}h^+_{n-1}}{h^+_{n-1}+h^-_{n-1}}       & 1+ z_{n-1}                    \\
                  \end{array}\right) \\[15pt]
 \left(\begin{array}{@{}c@{}}
p_0 \\
p_1     \\
p_2     \\
p_3     \\
\vdots \\
p_{n-3}       \\
p_{n-2}       \\
 p_{n-1}    \\
                  \end{array}\right) \\[15pt]=
 \left(\begin{array}{@{}c@{}}
0 \\
0     \\
0       \\
0       \\
\vdots \\
0    \\
 0    \\
 \frac{z_{n-1} h^-_{n-1}}{h^+_{n-1}+h^-_{n-1}} ATM  \\
                  \end{array}\right) \\[15pt].
\]
%0           & -z_2      & 1+2z_2 &  -z_2   & \cdots  & 0             & 0                  & 0         	                      \\
The last row of the later gives a further condition:
\begin{eqnarray}
-z_{n-1}h^+_{n-1}p_{n-2} + (1+z_{n-1})p_{n-1}(h^+_{n-1}+h^-_{n-1}) = z_{n-1}h^-_{n-1}ATM.
\label{putc}
\end{eqnarray}
The same approach can be repeated for all OTM calls $c_{n+1}, \dots\ c_{n+m}$ together with ATM call $c_n=ATM$ and the first ITM call $c_{n-1}$. 
Working in the same vein with~(\ref{tridc}) we use the first equation of the (reduced) call based system to get:
\begin{eqnarray}
-z_{n+1}h^-_{n+1}c_{n+2} + (1+z_{n+1})c_{n+1}(h^+_{n+1}+h^-_{n+1}) = z_{n+1}h^+_{n+1}ATM.
\label{callc}
\end{eqnarray}
The ATM term $z_n$ of the put~(\ref{tridp}) or the call~(\ref{tridc}) systems,  is particularly simple in the shifted SABR approach:
\begin{eqnarray}
z_n & = & \frac{T\theta(F)^2}{h^+_nh^-_n},\nonumber\\
      & = & \frac{T\vartheta(F)^2\kappa(F)}{h^+_nh^-_n},\nonumber\\
      & = & \frac{2T}{h^+_nh^-_n} \alpha^2 (F+b)^{2\beta}.
\end{eqnarray}
We use it together with~(\ref{atmterm}) to derive an approximation for $\alpha$
\begin{eqnarray}
\alpha &=& \frac{1}{(F+b)^{\beta}}\sqrt{\frac{h^+_nh^-_n}{2T}}\sqrt{\frac{ATM(h^+_n+h^-_n) }{p_{n-1}h^-_n + p_{n+1}h^+_n - ATM(h^+_n+h^-_n)}}.
\end{eqnarray}
Two equations~(\ref{putc}) and~(\ref{callc})
can be used to imply parameters $\rho$ and $\nu$ from option prices close ATM. More explicitly, we have:
\begin{eqnarray}
z_{n-1} &=&  \frac{p_{n-1}(h^+_{n-1}+h^-_{n-1})}{ p_{n-2}h^+_{n-1} +ATMh^-_{n-1} - p_{n-1}(h^+_{n-1}+h^-_{n-1}) },\nonumber\\
z_{n+1}&=&  \frac{c_{n+1}(h^+_{n+1}+h^-_{n+1})}{ c_{n+2}h^-_{n+1} + ATMh^+_{n+1} - c_{n+1}(h^+_{n+1}+h^-_{n+1})}.
\end{eqnarray}
with
\begin{eqnarray}
\label{zi}
z_i &=& \frac{ T\vartheta(k_i)^2\kappa(k_i)}{h^+_ih^-_i},\nonumber\\
\vartheta(k) &=& \alpha J(y) (k+b)^{\beta},\\
\label{ktoy}
y(k)&=&\frac{1}{\alpha}\int^F_k (u+b)^{-\beta}du,\\
J(y) & = & \sqrt{1 - 2\rho \nu y + \nu^2 y^2}.
\end{eqnarray}
Moving all the terms in~(\ref{zi}) independent of $\nu$ and $\rho$ to one side we obtain a system:
\begin{eqnarray}
\label{nrm}
\nu^2 y(k_{n-1}) - 2\rho\nu  = \frac{1}{y(k_{n-1})}\left(\frac{z_{n-1}h^+_{n-1}h^-_{n-1}}{T \kappa(k_{n-1})\alpha^2   (k_{n-1} + b)^{2\beta}}-1\right),\\
\nu^2 y(k_{n+1}) - 2\rho\nu  =  \frac{1}{y(k_{n+1})}\left(\frac{z_{n+1}h^+_{n+1}h^-_{n+1}}{T \kappa(k_{n+1})\alpha^2   (k_{n+1} + b)^{2\beta}}-1\right).
\label{nrp}
\end{eqnarray}
This system can be solved analytically for $\nu$ and $\rho$. We obtain the following result:
\begin{Th}
\label{th1}
In Andreasen-Huge one time step approximation for the shifted SABR SDE, the parameters $\alpha$, $\nu$ and $\rho$ have the following analytic expressions:
\begin{eqnarray}
\label{alpha}
\alpha &=& \frac{1}{(F+b)^{\beta}}\sqrt{\frac{h^+_nh^-_n}{2T}}\sqrt{\frac{ATM(h^+_n+h^-_n) }{p_{n-1}h^-_n + p_{n+1}h^+_n - ATM(h^+_n+h^-_n)}},\\
\label{nu2}
\nu^2 & = &\frac{\frac{z_{n-1}h^+_{n-1}h^-_{n-1}}{ y(F-h^-_n)  (F - h^-_n + b)^{2\beta}\kappa_{h^-_n}}
- \frac{z_{n+1}h^+_{n+1}h^-_{n+1}}{y(F+h^+_n)(F+h^+_n + b)^{2\beta} \kappa_{h^+_n} }}{T\alpha^2(y(F-h^-_n) - y(F+h^+_n))}
 - \frac{1}{y(F-h^-_n) y(F+h^+_n)},\\
\rho &=& \frac{1}{2\nu}\left(\frac{\frac{y(F+h^+_n)z_{n-1}h^+_{n-1}h^-_{n-1}}{ y(F-h^-_n)  (F - h^-_n + b)^{2\beta}\kappa_{h^-_n}}
- \frac{y(F-h^-_n)z_{n+1}h^+_{n+1}h^-_{n+1}}{  y(F+h^+_n)(F+h^+_n + b)^{2\beta}\kappa_{h^+_n}}}{T\alpha^2(y(F-h^-_n) - y(F+h^+_n))} - \frac{y(F-h^-_n) - y(F+h^+_n)}{y(F-h^-_n) y(F+h^+_n)}\right),
\label{rho} 
\end{eqnarray}
where 
\begin{eqnarray}
z_{n-1} &=&  \frac{p_{n-1}(h^+_{n-1}+h^-_{n-1})}{ p_{n-2}h^+_{n-1} +ATMh^-_{n-1} - p_{n-1}(h^+_{n-1}+h^-_{n-1}) },\nonumber\\
z_{n+1}&=&  \frac{c_{n+1}(h^+_{n+1}+h^-_{n+1})}{ c_{n+2}h^-_{n+1} + ATMh^+_{n+1} - c_{n+1}(h^+_{n+1}+h^-_{n+1})},\nonumber\\
\kappa_h &=& 1 -  \frac{h}{\sigma\phi(h/\sigma)}\left(\frac{1}{2} - \int^{h/\sigma}_0\phi(s)ds\right).
\end{eqnarray}
\end{Th}
For a uniform grid all the steps $h^{\pm}_j$ are equal and the theorem simplifies to the result quoted in the introduction.

It is interesting to compare the analytic result of Theorem~\ref{th1} with the classical calibration approach to imply SSABR parameters. For example, parameter alpha can be implied approximately as a solution to a cubic~\cite{W} whose coefficients depend on the ATM option value. 
The solution is not unique, and the choice of a sensible root may present a challenge. On the other hand, Theorem~\ref{th1} gives a unique explicit solution using three option prices.    

\section{Limiting behavior}

In this section we study the limiting behavior of~(\ref{alpha}-\ref{rho}) as the grid step $h^{\pm}_n$ goes to zero. Without lose of generality, here, we can assume that the grid is uniform.  First, we derive an analytic expression for $\alpha$:
\begin{eqnarray}
\alpha &=&  \lim_{h\to 0} \frac{h}{(F+b)^{\beta}}\sqrt{\frac{1}{T}}\sqrt{\frac{ATM }{p_{n-1} + p_{n+1} - 2ATM}},\\
&=& \frac{1}{(F+b)^{\beta}}\sqrt{\frac{1}{T}}\sqrt{\frac{ATM }{pdf_{ATM}}}.
\label{alim}
\end{eqnarray}
If we assume that the forward swap rate PDF is approximately Gaussian normal we can recover a popular approximation:
\begin{eqnarray}
\alpha &\approx&\frac{1}{(F+b)^{\beta}}\sqrt{\frac{1}{T}}\sqrt{\sigma_{ATM}^2 T } = \frac{\sigma_{ATM}}{(F+b)^{\beta}}.
\end{eqnarray}
Next, we look at the term $\rho\nu$ in~(\ref{nrm},\ref{nrp}) .  We shall be using the following approximations:
\begin{eqnarray}
z_{n-1}h^2 &=& \frac{p_{n-1}h^2}{p_{n-2} + ATM - 2 p_{n-1}} \approx \frac{p(F-h)}{pdf(F-h)},\\  
z_{n+1}h^2 &=& \frac{c_{n+1}h^2}{c_{n+2} + ATM - 2 c_{n+1}} \approx \frac{p(F+h) - h }{pdf(F+h)}.\\  
\end{eqnarray}
The transformation~(\ref{ktoy}) is a change of coordinates with a Jacobian:
\begin{eqnarray}
\frac{\partial y(k)}{\partial k} &=& - \frac{1}{\alpha(k+b)^{\beta}}, \quad{} y(F) = 0.
\end{eqnarray}
We use this change of coordinates to factor out a derivative with respect to $y(k)$ first from~(\ref{nrm}) by using~(\ref{alim}):
\begin{eqnarray}
\rho\nu &=& \frac{1}{2}\lim_{h\to 0} \frac{1}{y(k_{n-1})}\left(1-\frac{2z_{n-1}h^2}{T \kappa(k_{n-1})\alpha^2   (k_{n-1} + b)^{2\beta}}\right),\\
&=&\frac{1}{T\alpha^2}\lim_{h\to 0} \frac{1}{y(k_{n-1})}\left(\frac{p(F)}{\kappa(F)pdf(F)  (F + b)^{2\beta}} - \frac{p(F-h)}{\kappa(F-h)pdf(F-h)  (F-h + b)^{2\beta}}\right),\nonumber\\
&=&-\frac{1}{T\alpha^2}\frac{\partial_-}{\partial y}\frac{p(k(y))}{\kappa(k(y))pdf(k(y))  (k(y) + b)^{2\beta}}|_{k(y) =F},
\end{eqnarray}
where by $\partial_-$ we denoted the left derivative with respect to $y$.
Similarly, using~(\ref{nrp}) we get
\begin{eqnarray}
\rho\nu &=&-\frac{1}{T\alpha^2}\frac{\partial_+}{\partial y}\frac{c(k(y))}{\kappa(k(y))pdf(k(y))  (k(y) + b)^{2\beta}}|_{k(y) =F},
\end{eqnarray}
with $\partial_+$ being the right derivative with respect to $y$. We can glue the functions
\begin{eqnarray}
-\frac{1}{T\alpha^2}\frac{\partial_-}{\partial y}\frac{p(k(y))}{\kappa(k(y))pdf(k(y))  (k(y) + b)^{2\beta}}&& -\frac{1}{T\alpha^2}\frac{\partial_+}{\partial y}\frac{c(k(y))}{\kappa(k(y))pdf(k(y))  (k(y) + b)^{2\beta}}
\end{eqnarray}
at $k(y)=F$.  Using~(\ref{nu2}) to see that the result is a differentiable function (as $y(k_{n-1})$ and $y(k_{n+1})$ are independent), we arrive at:  
\begin{eqnarray}
\label{sqbr}
\nu^2 & = &\frac{2}{T\alpha^2}\frac{\partial^2_-}{\partial y^2}\frac{p(k(y))}{\kappa(k(y))pdf(k(y))  (k(y) + b)^{2\beta}}|_{k=F}\\
&=&\frac{2}{T\alpha^2}\frac{\partial^2_+}{\partial y^2}\frac{c(k(y))}{\kappa(k(y))pdf(k(y))  (k(y) + b)^{2\beta}}|_{k=F}\\
\end{eqnarray}
We summarise our findings in  the next theorem:
\begin{Th}
In the short maturity approximation of Andreasen-Huge SSABR model the following analytic expressions for model parameters $\alpha$, $\nu$ and $\rho$ hold:
\begin{eqnarray}
\alpha &=& \frac{1}{(F+b)^{\beta}}\sqrt{\frac{1}{T}}\sqrt{\frac{ATM }{pdf_{ATM}}},
\end{eqnarray}
\begin{eqnarray}
\nu= \frac{1}{\alpha}\sqrt{\frac{2}{T}}\sqrt{\frac{\partial^2_-}{\partial y^2}\frac{p(k)}{\kappa(k)pdf(k)  (k + b)^{2\beta}}}|_{k=F}& =& \frac{1}{\alpha}\sqrt{\frac{2}{T}}\sqrt{\frac{\partial^2_+}{\partial y^2}\frac{c(k)}{\kappa(k)pdf(k)  (k + b)^{2\beta}}}|_{k=F},\nonumber\\
&&\\
\rho =- \frac{1}{\alpha\sqrt{2T}}\frac{\frac{\partial_-}{\partial y}\frac{p(k)}{\kappa(k)pdf(k)  (k + b)^{2\beta}}|_{k=F}}{\sqrt{\frac{\partial^2_-}{\partial y^2}\frac{p(k)}{\kappa(k)pdf(k)  (k + b)^{2\beta}}}|_{k=F}} &=& - \frac{1}{\alpha\sqrt{2T}}\frac{\frac{\partial_+}{\partial y}\frac{c(k)}{\kappa(k)pdf(k)  (k + b)^{2\beta}}|_{k=F}}{\sqrt{\frac{\partial^2_+}{\partial y^2}\frac{c(k)}{\kappa(k)pdf(k)  (k + b)^{2\beta}}}|_{k=F}},
\end{eqnarray}
where 
\begin{eqnarray}
y(k)&=&\frac{1}{\alpha}\int^F_k (u+b)^{-\beta}du.
\end{eqnarray}
\end{Th}

\section{Parameter recalibration}
In this section we show how analytic formulae from Section 3 can be used for recalibration of different versions of SABR to each other.

First, we consider  an example of EUR 10y 10y Swaption and demonstrate how the  analytic skew calibration improves matching between Hagan and Andreasen-Huge SSABR approximations.
We use Hagan SSABR parameters: 
\begin{equation}
b = 3\%, \quad \beta = 40\%, \quad \alpha = 2.17\%, \quad \rho = -23.78\%, \quad \nu = 26.12\%
\end{equation}
Using formulae from Section 3 we find calibrated skew parameters to use in Andreasen-Huge SSABR as
\begin{equation}
b = 3\%, \quad \beta = 40\%, \quad \alpha = 2.06\%,  \quad \rho = -26.84\%, \quad \nu = 27.54\%.
\end{equation}
The comparison is plotted in Figure~1.
\begin{figure}[h!]
\label{eur10y10y}
\centering
\includegraphics[height=75mm]{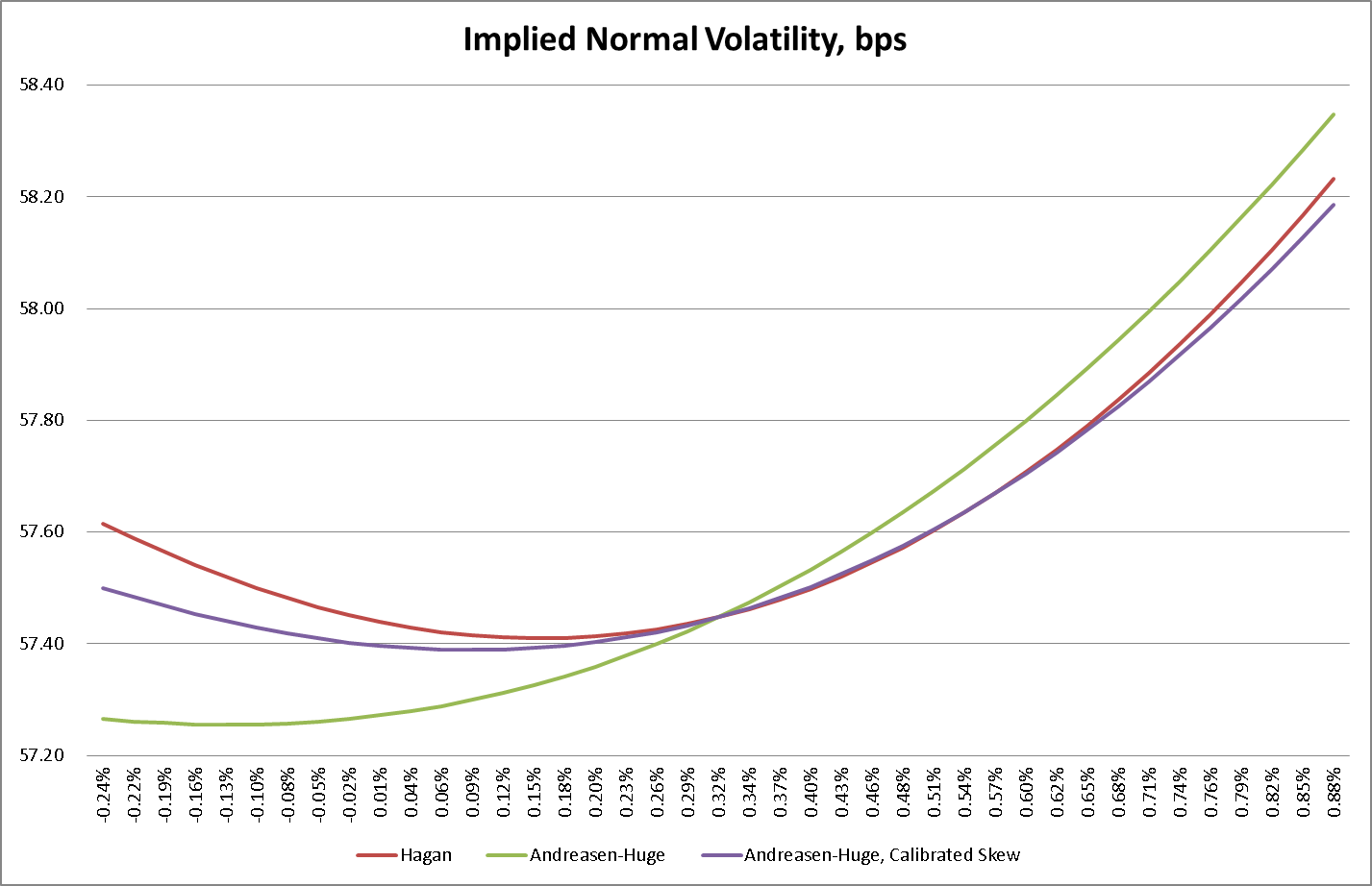}
\caption{\footnotesize{EUR 10y 10y Swaption Skew.}}
\end{figure}

Next we show how to recalibrate Andreasen Huge SSABR following a change of the parameter beta. We use the same Hagan data as in the previous example, but change beta from 40\% to 60\%. The recalibrated skew parameters for Andreasen-Huge SSABR are:
\begin{equation}
b = 3\%, \quad \beta = 60\%, \quad \alpha =4.08\%,  \quad \rho = -35.88\%, \quad \nu = 29.50\%
\end{equation}
The comparison is plotted in Figure~2.
\begin{figure}[h!]
\label{eur10y10ybeta60}
\centering
\includegraphics[height=75mm]{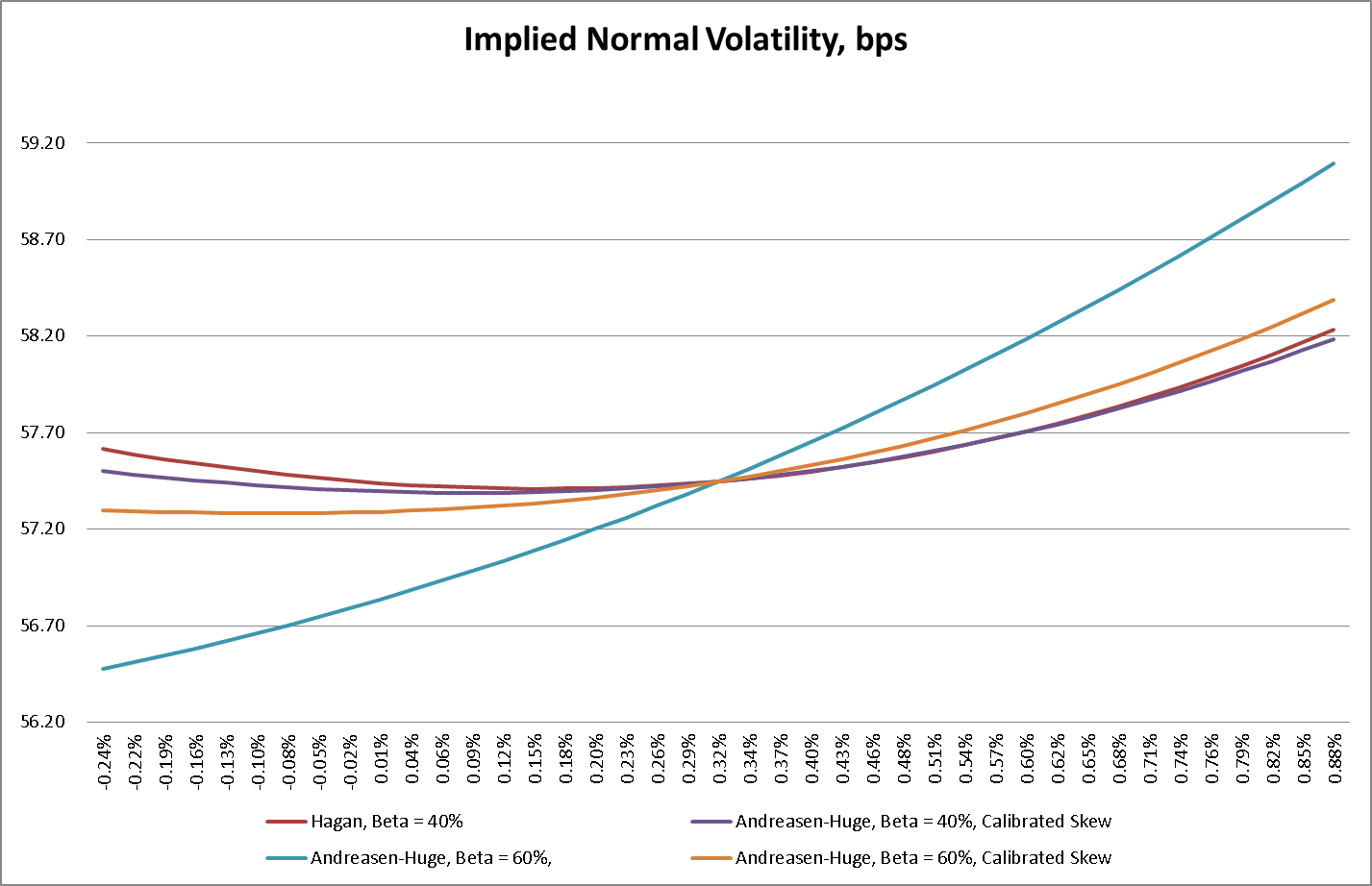}
\caption{\footnotesize{EUR 10y 10y Swaption Skew, Hagan beta 40\% vs Andreasen-Huge beta 60\%.}}
\end{figure}

Similarly we consider beta change from 40\% to 20\%. The recalibrated skew parameters for Andreasen-Huge SSABR are:
\begin{equation}
b = 3\%, \quad \beta = 20\%, \quad \alpha =1.05\%,  \quad \rho = -16.27\%, \quad \nu =25.84\%
\end{equation}
The comparison is plotted in Figure~3.
\begin{figure}[h!]
\label{eur10y10ybeta20}
\centering
\includegraphics[height=70mm]{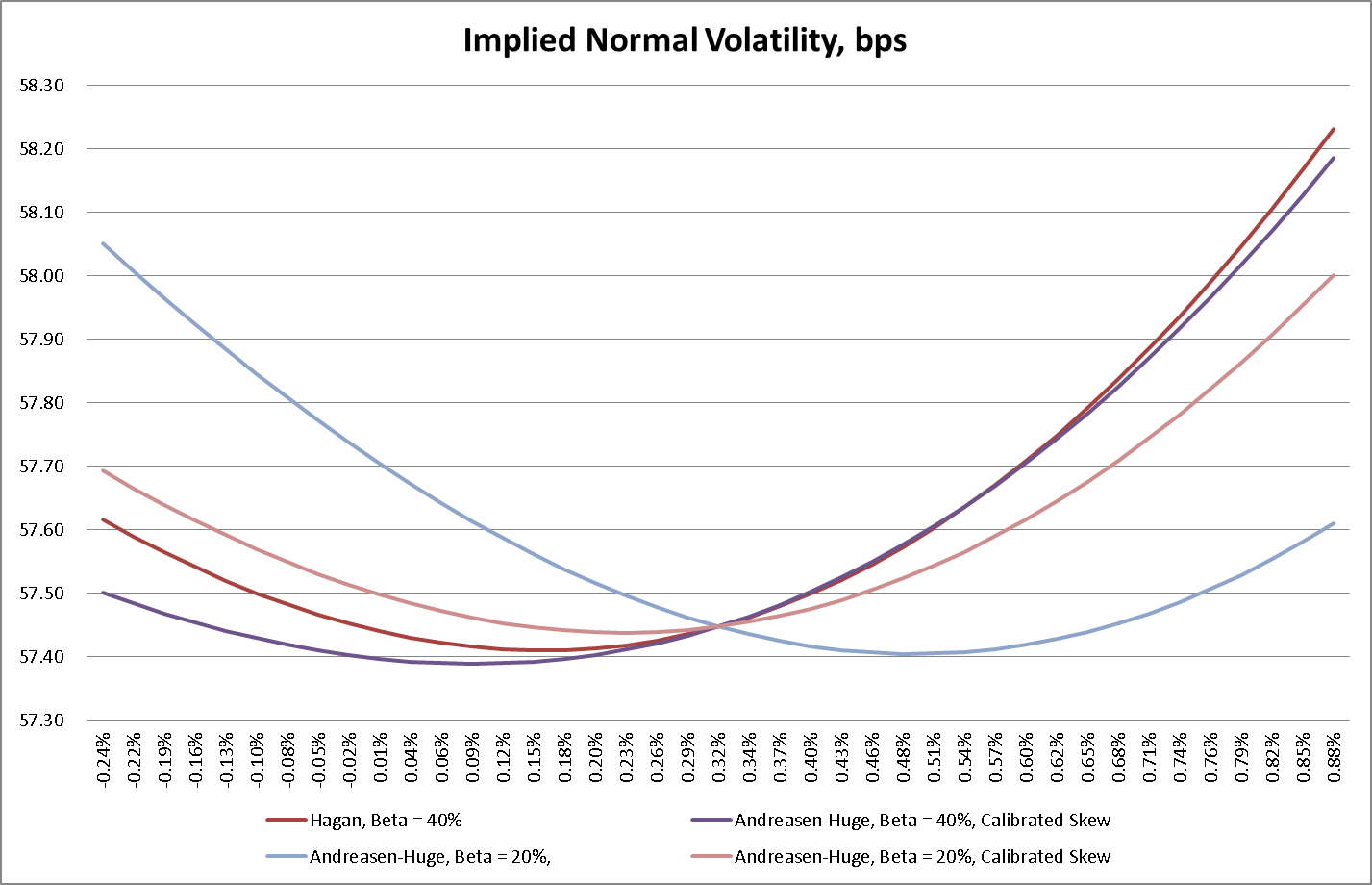}
\caption{\footnotesize{EUR 10y 10y Swaption Skew, Hagan beta 40\% vs Andreasen-Huge beta 20\%.}}
\end{figure}

We shall now consider a change of shift from 3\% to 3.25\%. The recalibrated skew parameters for Andreasen-Huge SSABR are:
\begin{equation}
b = 3.25\%, \quad \beta = 40\%, \quad \alpha =2.01\%,  \quad \rho = -25.57\%, \quad \nu =27.17\%
\end{equation}
The comparison is plotted in Figure~4.
\begin{figure}[h!]
\label{eur10y10yshift3.25}
\centering
\includegraphics[height=70mm]{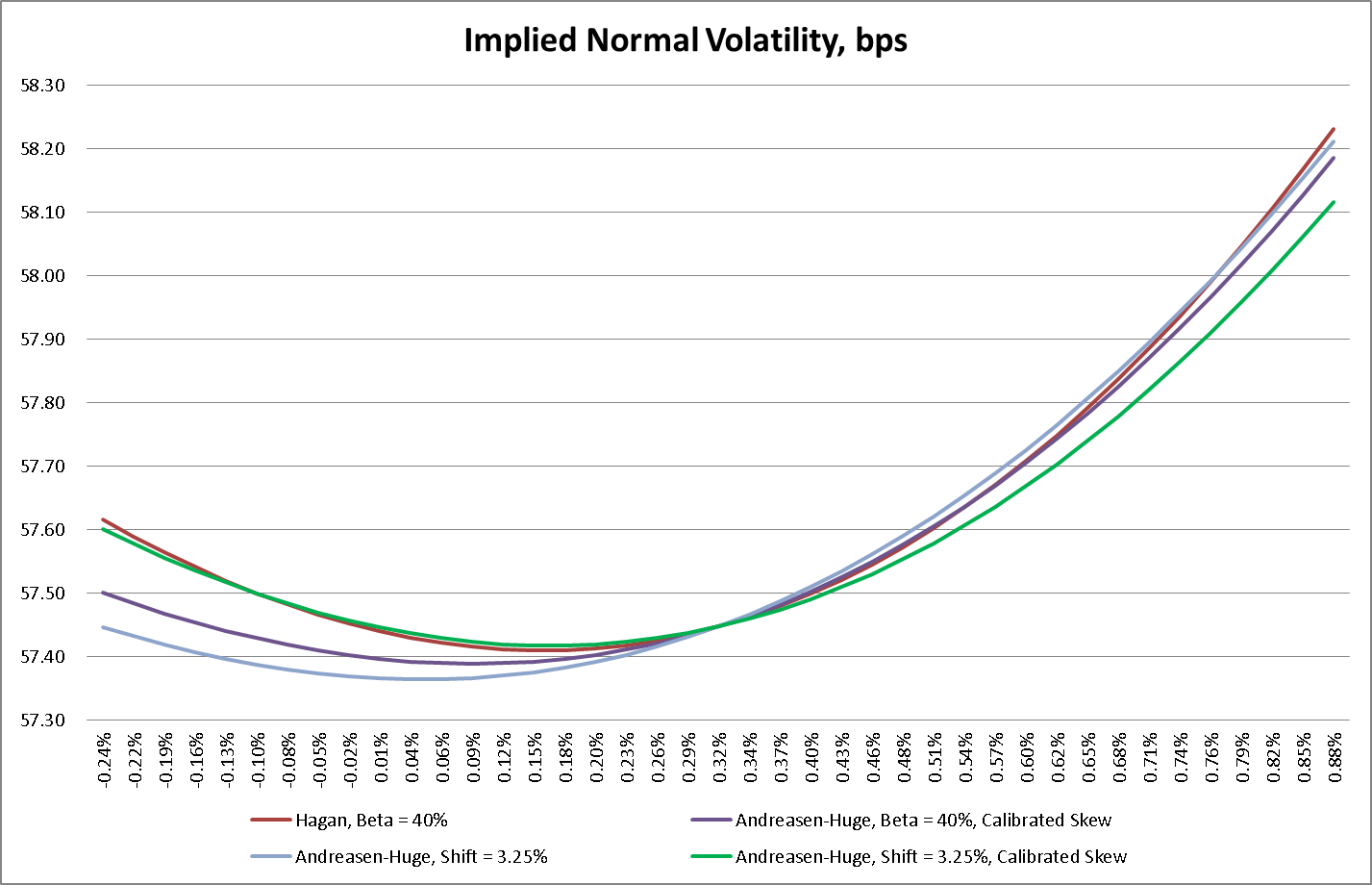}
\caption{\footnotesize{EUR 10y 10y Swaption Skew, Hagan shift 3\% vs Andreasen-Huge shift 3.25\%.}}
\end{figure}

If we also change Andreasen-Huge beta from 40\% to 50\% we obtain recalibrated skew parameters for Andreasen-Huge SSABR as:
\begin{equation}
b = 3.25\%, \quad \beta = 50\%, \quad \alpha =2.80\%,  \quad \rho = -30.10\%, \quad \nu =28.01\%
\end{equation}
The comparison is plotted in Figure~5.
\begin{figure}[h!]
\label{eur10y10ybeta50}
\centering
\includegraphics[height=70mm]{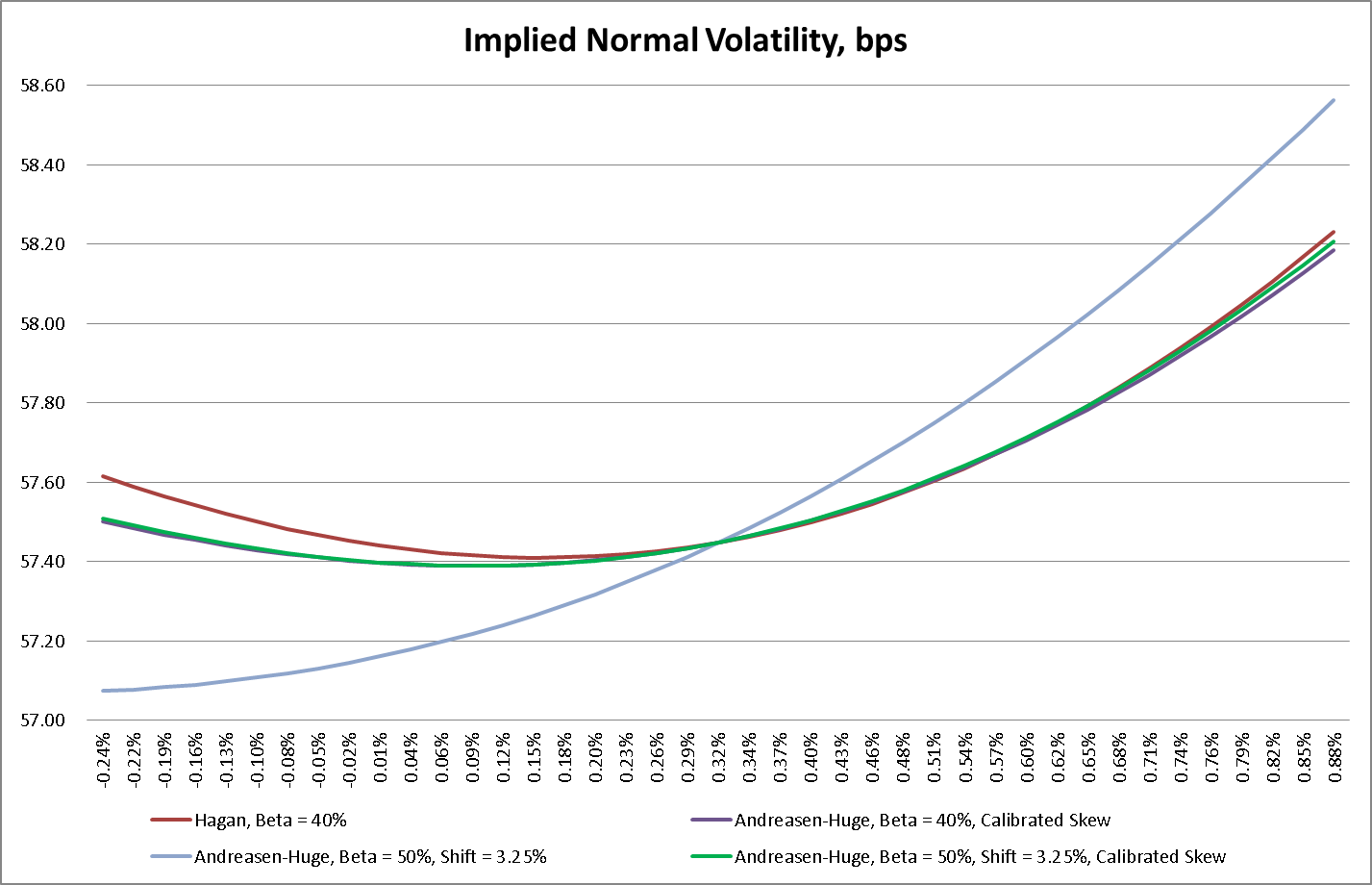}
\caption{\footnotesize{EUR 10y 10y Swaption Skew, Hagan shift 3\%, beta 40\% vs Andreasen-Huge shift 3.25\%, beta 50\%.}}
\end{figure}

While analytic skew calibration allows us to make a closer match between Hagan and Andreasen-Huge SSABR implementations, we cannot achieve a perfect match between the two due to numerical limitations. Firstly, the calibration only operates in a small neighborhood of ATM and there is an unavoidable divergence at the wings.
Secondly, the use of various numerical procedures to interpolate and extrapolate Andreasen-Huge SSABR in to non-grid points leads to  numerical noise that prevents us from having an exact comparison.

\section{IR Futures Options}

One of the markets where the results of this paper can be applied directly is the IR futures market. The IR futures options market quotes are available for a wide range of equally spaced strikes, where the distance between consecutive strikes is 12.5bps - this is a typical grid size for a sensible (uniform) Andreasen-Huge SABR grid.  We test the results of the paper on Eurodollar options contracts. 

We looked at EDH3P and EDH3C Eurodollar options on 04/01/2021:
\begin{figure}[h!]
\label{EDH3P}
\centering
\includegraphics[height=25mm]{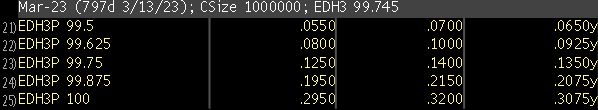}
\caption{\footnotesize{Eurodollar put options expiring March 2023. Quotes as of 04/01/2021.}}
\end{figure}
\begin{figure}[h!]
\label{EDH3C}
\centering
\includegraphics[height=25mm]{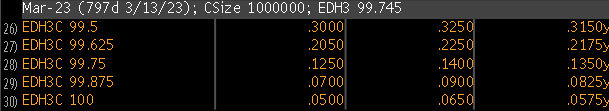}
\caption{\footnotesize{Eurodollar call options expiring March 2023. Quotes as of 04/01/2021.}}
\end{figure}

Using quoted ATM last price (0.1350y) and OTM last prices (last columns of the figures above) we obtained the follwoing values for SSABR parameters on a uniform grid of 241 points between $-5\%$ and $+25\%$:
\begin{equation}
\beta= 5\% \quad  fixed; \quad \alpha= 0.2079\% \quad   analytic; \quad \nu= 108.62\% \quad  analytic; \quad  \rho= 35.71\%\quad  analytic; \\
\end{equation}

Using these parameters we calculated the implied volatility by Andreasen-Huge SSABR and compared it to the implied volatility quoted on the Eurodollar market for options on EDH3 Future:
\begin{figure}[h!]
\label{EDH3Vol}
\centering
\includegraphics[height=60mm]{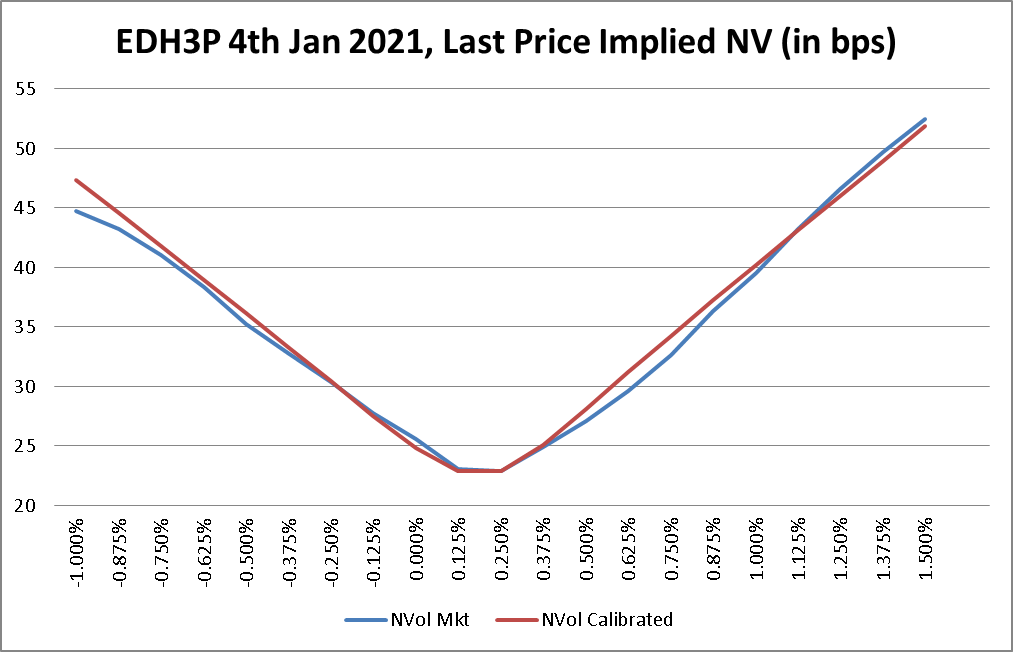}
\caption{\footnotesize{Eurodollar put options expiring March 2023. Implied BP Normal Vol as of 04/01/2021.}}
\end{figure}

\section*{Conclusion}
We derived analytic formulae which express SSABR parameters $\alpha$, $\nu$ and $\rho$ in Andreasen-Huge one time step framework in terms of five option prices close ATM. We used these formulae to give a characterisation for SSABR parameters in terms of swap rate forward distribution. 

We showed how the formulae can be used for calibrating Andreasen-Huge SSABR model to analytical approximation for the solution of the SSABR SDE, as well as how Andreasen-Huge SSABR model can be recalibrated after changes of shift or beta parameters.
The later procedure can be used for constructing consistent risk in the multi parametric SSABR model where different regions of the strike space are governed by different SABR regimes.   

One can even hide SABR parameters completely from the user's eye and run their volatility model in terms of five options' prices specified at equally spaced strikes around ATM.  The analytic expressions for SSABR parameters allow hedging of the IR skew and smile done exactly in terms of five options which are liquidly available in the Bond and Eurodollar futures markets.

\end{document}